\pdfoutput=1
\documentclass[10pt,conference]{IEEEtran}
\makeatletter
\def\ps@headings{%
\def\@oddhead{\mbox{}\scriptsize\rightmark \hfil}%
\def\@evenhead{\scriptsize\thepage \hfil \leftmark\mbox{}}%
\def\@oddfoot{}%
\def\@evenfoot{}}
\makeatother

\usepackage{fancyhdr}
\usepackage{graphicx, subfigure}
\usepackage{verbatim}
\usepackage{verbatimbox}

\usepackage{algorithm}
\usepackage{algorithmic}
\usepackage{array}

\usepackage[geometry]{ifsym}
\usepackage{array}
\usepackage{mathtools}
\usepackage{mathptmx} 
\usepackage{xspace}
\usepackage{color}
\usepackage{url}

\usepackage{perpage}	 
\MakePerPage{footnote}	 

\usepackage[hang,splitrule]{footmisc}

\def\name{FIXME\xspace}
\newcommand{\cut}[1]{}

\usepackage[font=small,labelfont=bf]{caption}



\renewcommand{\footnoterule}{%
	\kern -3pt
	\hrule width 2in
	\kern 2.6pt
}

\newenvironment{noitemize}{
 \begin{list}{{\textbullet}}{
  \setlength{\partopsep}{0pt}
  \setlength{\parskip}{0pt}
  \setlength{\parsep}{0pt}
  \setlength{\topsep}{0pt}
  \setlength{\itemsep}{0pt}
  \setlength{\itemindent}{0pt}
  \setlength{\leftmargin}{9pt}
 }
}{
 \end{list}
}

\begin{document}

\date{}

\title{\name: Enhance Software Reliability with Hybrid Approaches in Cloud}

\author{\IEEEauthorblockN{Jinho Hwang, Larisa Shwartz, Qing Wang, Raghav Batta}
\IEEEauthorblockA{\textit{IBM T.J. Watson Research Center} \\
Yorktown Heights, NY, USA\\
\{jinho, lshwart\}@us.ibm.com, \{qing.wang1, raghav.batta1\}@ibm.com}
\and
\IEEEauthorblockN{Harshit Kumar}
\IEEEauthorblockA{\textit{IBM Research India} \\
New Delhi, India \\
harshitk@in.ibm.com}
\and
\IEEEauthorblockN{Michael Nidd}
\IEEEauthorblockA{\textit{IBM Research Europe} \\
Zurich, Switzerland \\
mni@zurich.ibm.com}
}

\maketitle

\begin{abstract}

With the promise of reliability in cloud, more enterprises are migrating to
cloud.  The process of continuous integration/deployment (CICD) in cloud
connects developers who need to deliver value faster and more transparently
with site reliability engineers (SREs) who need to manage applications
reliably.  SREs feed back development issues to developers, and developers
commit fixes and trigger CICD to redeploy. The release cycle is more continuous
than ever, thus the code to production is faster and more automated.  To
provide this higher level agility, the cloud platforms become more complex in
the face of flexibility with deeper layers of virtualization.  However,
reliability does not come for free with all these complexities.  Software
engineers and SREs need to deal with wider information spectrum from
virtualized layers.  Therefore, providing correlated information with true
positive evidences is critical to identify the root cause of issues quickly in
order to reduce mean time to recover (MTTR), performance metrics for SREs.
Similarity, knowledge, or statistics driven approaches have been effective, but
with increasing data volume and types, an individual approach is limited to
correlate semantic relations of different data sources.  In this paper, we
introduce \name to enhance software reliability with \emph{hybrid} diagnosis
approaches for enterprises.  Our evaluation results show using hybrid diagnosis
approach is about 17\% better in precision.  The results are helpful for both
practitioners and researchers to develop hybrid diagnosis in the highly dynamic
cloud environment. 


\end{abstract}


\begin{IEEEkeywords}
event management, hybrid system, event correlation, localization, cloud
\end{IEEEkeywords}

%

\section{Introduction}
\label{sec:introduction}


With the promise of reliability, cloud has become more flexible and dynamic to
provide continuous software development and deployment.  While the contemporary
microservices architecture has simplified the scope of software developers
through well defined representational state transfer application programming
interfaces (REST APIs), roles of site reliability engineers (SREs) towards
availability, latency, performance, efficiency, change management, monitoring,
emergency response, and capacity planning have become even more complex.  
Container deployments are more dynamic than ever, with lifespans of 10 seconds
or less becoming increasingly prevalent, emphasizing the need for real-time
visibility that delivers detailed audit and forensics
records~\cite{sysdig_report_2019}.  The ephemeral and immutable nature of
containers is advantageous for development and operations (DevOps), but
simultaneously can be challenging for software developers and SREs to correctly
diagnose incidents and resolve them timely.  Based on 2020 SRE report, 80\% of
SREs work on post-mortem analysis of incidents due to lack of provided
information and 16\% of toil come from investigating false
positives/negatives~\cite{sre_report_2020,10.1109/ICSE-SEIP.2019.00020}.

\begin{figure}[t]
\center
\includegraphics[bb=0 0 100 100, trim=0 0 0 0, clip, width=0.48\textwidth]{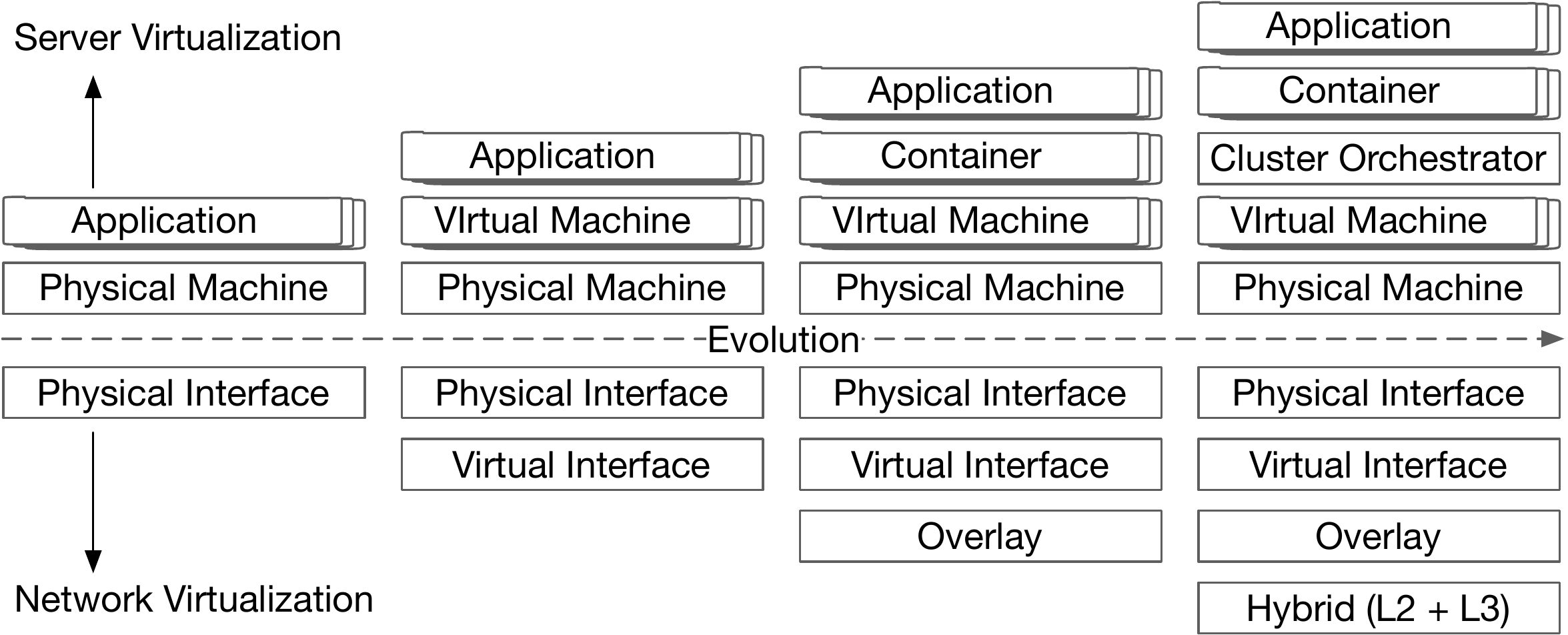}
\caption{Evolution of server and network virtualization}
\vspace{-4mm}
\label{fig:virtualization}
\end{figure}

As shown in Figure~\ref{fig:virtualization}, cloud has adopted more
virtualization technologies, in turn virtualized layers stack up to run
applications and the number of applications running in one node increases.
This also means that any virtualized layer below applications can have direct
impact on running applications.  This increased scope of application impact is
not only derived from server virtualization, but also from network
virtualization.  A microservices architecture has driven this as the number of
containers is more than the same type of a monolithic application.  That is,
more information such as logs, alerts, metrics are generated, thus
consolidating these information in a meaningful way becomes intractable.  As
noted in~\cite{sre_report_2020}$-$41\% of SREs have answered half or more of
their work is toil$-$, when dealing with the high amounts of toil in an
organization, the underlying reasons are lack of intelligence and automation.
To cope with this hardship, artificial intelligence for IT operations (AIOps)
has emerged to help SREs to recognize serious issues faster and with greater
accuracy than humans.  The ultimate objective of AIOps is to minimize mean time
to recover (MTTR) by providing software engineers and SREs with localized and
correlated information.  MTTR spans different stages including
mean-time-to-detect (MTTD), -identify (MTTI), -know (MTTK), -repair
(MTTRepair), and -resolve (MTTResolve).  From the large amount of data, the
problem determination and information correlation are the keys to start with
the right problem and correct information. Not to mention that this reduces the
number of false positives/negatives.

From various data sources such as logs, alerts, metrics, anomaly events, and
the like, the premise of information consolidation is that information have
some common elements to each other, appear within a time window, and occur in a
proximity of location.  The two main bodies of research include event
correlation~\cite{Mirheidari_2013} and problem
localization~\cite{10.1109/COMST.2016.2570599,fault_localization_survey}.  In
large, similarity-based, knowledge-based, and statistical approaches have been
used to identify patterns and groups, filter and prioritize the
events~\cite{Mirheidari_2013,10.1109/COMST.2016.2570599} for both pre-mortem
and post-mortem analysis.  However, in AIOps, not one approach can always
perform better than others because it is extremely hard to identify  
underlying patterns of different data sources of all levels and also
generalize knowledge obtained from one application to another. In practice, a
hybrid or ensemble approach would usually result in better correlation
results~\cite{43438,doi:10.1002/sec.1190,10.1016/j.comnet.2011.03.005}. 

In this paper, we introduce hybrid methodologies for AIOps used in production,
and evaluate how they improve problem determination and information
correlation. Our contributions include the following:

\begin{noitemize}
\item Explore various data sources in cloud native environment and their normalization,
\item Find patterns defined by complex predicates in large, constantly changing datasets,
\item Introduce hybrid diagnosis approaches for entity resolution and event correlation,
\item Incorporate continuous learning with feedback, and
\item Evaluate the hybrid diagnosis approaches and the impact of the continuous learning.
\end{noitemize}

\cut{
The goal for the correlation engine is to identify known event patterns, group,
filter and prioritize the events to make the tickets more readable, and
ultimately, to take some of the load off the operators.

- to identify extraordinary situations, - to identify the root-cause of a
problem, - or to make predictions about the future and to discover trends.

A false positive is an event or alarm indicating an extraordinary situation,
even though there is no problem, e.g A false negative is an event indicating,
that the situation is normal, even though there is a problem, e.g.
}

\section{Background and Challenges}
\label{sec:background}


Monitoring system is an automated system that provides an effective and
reliable means of ensuring that anomalous behavior or degradation of the vital
signs of hybrid application is flagged as a problem candidate (monitoring
event) and sent for diagnosis and resolution. Events from hybrid environment
are consolidated in an enterprise event management system (EM). EM often
incorporates rules that define whether to create an incident record (ticket)
for IT problem reporting. In some cases, SREs analyze incoming events or
symptoms before deciding on creating a ticket. Tickets are collected by
Incident, Problem, and Change (IPC) system. Independently, the information is
also collected by myriad of tools that ingest data from various sources and
provide events of their own. 

Operations or SREs perform problem determination, diagnosis, and resolution
based on the symptoms information in the ticket. In the interviews conducted
with SREs, they have identified diagnosis as the most difficult task. The
majority of SREs have pointed out that given right diagnosis, they would be
able quickly to derive actions required to resolve the issue. Being able to
troubleshoot a problem and to arrive to a diagnosis is often considered to be
an innate skill ~\cite{SRE_book_2017}. Problem determination is a
labor-intensive process, and Operations/SREs use any help they could get from
analytics.  There has been a great deal of effort spent on developing
methodologies for specifying and reasoning about symptoms/signals provided
through monitoring of systems, be they hardware or software. 

While there is large body of techniques in existence, each technique is usually
focused on a single type of data (events or log anomalies or metric analysis).
Lately, a number of service providers have embarked on extending their existing
analytics to other data types.  The benefit of this approach is shortening time
to market through adaptation of well-researched  capabilities; one obvious
drawback is that performance of any methodology is optimized for specific data,
and it does not perform as well on different data types. In this paper we
present in-depth review of the operational data types and combinations of
algorithms for working with each data type to achieve a goal of identifying a
group of symptoms with a common root cause and localizing the problem.

Although the methodology  we describe is successfully used for identification
and linkage of symptoms from variety  of data sources, we are not able to
adjust for some changes in the architecture and monitoring strategy. It is
imperative to continue improving the models through continuous learning and
feedback. We point out that in our experience, SREs do not tend to provide an
extensive feedback and more often than not their feedback is limited to
thumbs-up/thumbs-down response. Learning from this feedback is a challenge if
the insight that they assess is complex. For example, a negative feedback
received for a grouping of symptoms is missing critical information about which
signals do not belong to the group.  In this paper we describe an approach to
using their minimal feedback efficiently.  

It is a widely accepted fact today that operational data poses additional
challenges in comparison to social data for example. In this paper we describe
various data types used in operations and challenges associated with it. To
name a few, some IT data is often created using templates by different
developers, so the content of data of the same type could vary drastically;
data that represents a symptom of a problem only makes sense in the context of
system/application configuration or resource, however this critical information
is often embedded in the text and not provided as first class item.

\section{Related Work}
\label{sec:related}


The main body of related researches is information (events or alerts)
correlation.  Prior arts for information correlation are categorized into
similarity based, knowledge based and statistical approaches. 
The similarity-based event correlation works on the premise that two events
that have similar root causes should be grouped together.  The base logic is to
compute similarity between the feature spaces of data and measure the
score~\cite{10.5555/872016.872176,10.1109/TDSC.2004.21,Elshoush2013}. For
example, a weighted sum of feature similarity of two data points is used to
measure the score~\cite{10.5555/645839.670734}.  One of
the most popular similarity based methods is clustering. The clustering divides
data into a number of groups such that data in the same group are more similar
than other data in other groups.  Klaus~\cite{991517} proposes an alert
clustering approach for identification of root causes that trigger the alert.
Vaarandi~\cite{1251233} clusters for log event data which helps one to detect
frequent patterns from logs, to build log profiles, and to identify anomalous
logs.

Experience based knowledge often becomes the source of intelligence or
automation, which can be turned into rules, templates, or scenarios.  Kabiri
et. al.~\cite{kabiri2007} propose a rule-based temporal alert correlation
system that uses an inference engine to aggregate redundant alerts and derive
correlation between alerts using a scenario-based knowledge base.
Klaus~\cite{10.1145/950191.950192} mines rules by learning from previous
events, and use these rules to cluster the new incoming events as they occur.
Dain et. al.~\cite{Dain2001BuildingSF} map each incoming event to one of the
manually prepared predefined scenarios or patterns, where each scenario
represents a sequence of actions.  Another type of the knowledge based approach
is expert systems. The expert systems aim at reproducing the performance of a
human expert.  The approaches under expert systems constitute building a
simulator that generates alerts~\cite{10.5555/890538} or
learns alert
scenarios~\cite{multisteps_cyber,10.5555/597917.597921,10.1007/3-540-36084-0_7,10.1016/j.inffus.2009.01.005,10.1016/j.comcom.2008.11.012}
on software applications to simulate the various faults, further models are
proposed for an event correlator~\cite{10.1145/586110.586144,
10.1145/996943.996947} that correlates related
events. 



Statistical traits find latent characteristics of data and join them in
meaningful ways towards the objective.  Many prior arts use machine learning
algorithms to find patterns of data points and the repetition
patterns~\cite{10.1016/j.istr.2005.07.001}.
Among them, Dain et. al.~\cite{Dain2002} mine scenarios from historical events,
and classify each incoming event into one of the candidate scenarios using
probabilistic methods.  Smith et.
al.~\cite{Smith2005ClusteringUA} use a
hierarchical unsupervised machine learning structure to identify the first and
second levels of groups that bubble up. They also use an auto-encoder to learn
the event distribution, and to correlate events together.  Pietraszek et.
al.~\cite{10.1016/j.istr.2005.07.001} propose an ensemble
approach to use various machine learning algorithms such as support vector
machines, decision trees and na\"ive bayes to suppress false positives.  Peter
et. al.~\cite{10.1109/PDP.2010.80} define activation patterns that store
information in an associative network graph and perform a graph analysis to
find common patterns based on nominal and distance based features.


To the best of our knowledge, we are the first to investigate the hybrid
diagnosis system that spans all of similarity based, knowledge based, and
statistical approaches for entity resolution and event correlation.

\section{Data}
\label{sec:data}

As a rule of thumb to build AI systems, \textit{results can only be as good as
quality of data}.  Understanding types of data and guaranteeing data quality
are the first steps towards a better AI system.  This section defines terminologies
and discusses details about input data.

\subsection{Terminology}

An {\bf event} indicates that something of note has happened and is associated
with one or more applications, services, or other managed resources. For instance,
a container has moved to new node, column added to a DB table, a new version of
an application is deployed, or memory or CPU exhausted. One or more events can
turn into any form of alerts or anomalies based on the deviation
from what is defined as standard, normal or expected.
Anomaly detection (a.k.a. outlier analysis) is a step in data mining that identifies
events and/or observations that deviate from a dataset's normal behavior.
Anomalous data can indicate critical incidents, such as a technical glitch,
or potential opportunities, for instance a change in consumer behavior.

An {\bf alert} is a record (type) of an event indicating a (fault) condition in
the managed environment. It requires or will require in the future, human or
automatic attention and actions toward remediation. For instance, disk drive
failure or network link down could be alerts.
An {\bf incident} represents a reduction in the quality of a business
application or service. It is driven by one or more alerts. Incidents require
prompt attention. For instance, application unresponsive or storage array
inaccessible could be serious outages.
A {\bf ticket} is an actionable or incident embodied in a service desk tool,
where the client has decided to use one.
A {\bf change record} is a description of a change that should be deployed,
including configuration changes, code changes, security updates, and so on.

\subsection{Logs / Metrics}
\label{subsec:logs_metrics}

Logs and metrics are two fundamental data sources generated from every level of
components as shown in Figure~\ref{fig:virtualization}.  A log is an event
happened and a metric is a measurement of the health of a system.  In general,
{\bf logs} include informational, debug, warning, error, and critical depending
on the severity.  In production systems, only warning, error and critical logs
may be collected.  In each log line, details about the event such as a resource
that was accessed, who accessed it, and the time are included.  Each log is
meant to have different sets of data so that the problem
localization can be obvious.

While logs are about a specific event, {\bf metrics} are a measurement at a
point in time.  Each metric data point can have value, timestamp, and identifier
of what that value applies to (like a source or a tag).  While logs may be
collected any time an event takes place, metrics are typically collected at
fixed-time intervals, thus called a time-series metric.  A sudden rise and spike
of CPU or memory utilization render alerts.  However, without logs, it is hard
to understand what causes such spikes.  Therefore, bringing together both logs
and metrics can provide much more clarity.  During spikes, there may be some
unusual log entries indicating the specific event that caused the spike.

From logs and metrics, SREs look for important signals that can lead to the
problem diagnosis and potentially resolution.  Broadly, four golden
signals\footnote{\url{https://landing.google.com/sre/sre-book/chapters/monitoring-distributed-systems}}
are known as the most helpful signals (events):

\noindent {\bf Latency}: Latency is the time it takes to service a request. While latency can be captured for both successful requests and failed requests, it is important to differentiate between them.
The failed requests may increase
the overall latency, but this may not necessarily be the latency of the system.
At the same time, a slow error is even worse than a fast error, so it is
important to track error latency, as opposed to just filtering out errors.

\noindent {\bf Traffic}: Traffic is a measure of demand for the system, measured
in a high-level system-specific metric.  For instance, too many HTTP requests to
a web server or API may result in additional stress on the system, triggering
downstream effects.  The traffic signal helps differentiate capacity problems
from improper system configurations that can cause problems even during low
traffic. 

\noindent {\bf Errors}: This is the rate of requests that fail, either
explicitly (e.g., HTTP 500s), implicitly (e.g., an HTTP 200 success response,
but coupled with the wrong content), or by policy (e.g., service level objective
(SLO)).  It is useful to diagnose misconfigurations in your infrastructure, bugs
in your application code, or broken dependencies.

\noindent {\bf Saturation}:
Saturation is the load on the system resources (e.g., CPU utilization, memory
usage, disk capacity, and operations per second).  Note that many systems
degrade in performance before they achieve 100\% utilization, so having a
utilization target is essential.  
While parts of the system become saturated first,
often, these metrics are leading indicators, so capacity can be adjusted before
performance degrades.

\subsection{Events / Alerts}

While the main sources of operational data are metrics and logs (\S
\ref{subsec:logs_metrics}), based on the rules or algorithms, they become
events. Other artifacts such as configuration, security, code changes, and the
like can be sources of events.  Anomalous---also, based on rules or
algorithms---events become alerts.  As shown in
Figure~\ref{fig:virtualization}, every component in virtualized layers can
produce events and parts of them can become alerts (i.e., important error
signals). In often cases, not a single alert can tell what went
wrong exactly, so correlating information with patterns would likely lead to
better diagnosis.  For example, a spike in error rate could indicate the
failure of a database or network outage. Also, following a code deployment, it
could indicate bugs in the code that somehow survived testing or only surfaced
in the production environment.




\begin{figure}[t]
\center
\includegraphics[bb=0 0 100 100, trim=0 0 0 0, clip, width=0.45\textwidth]{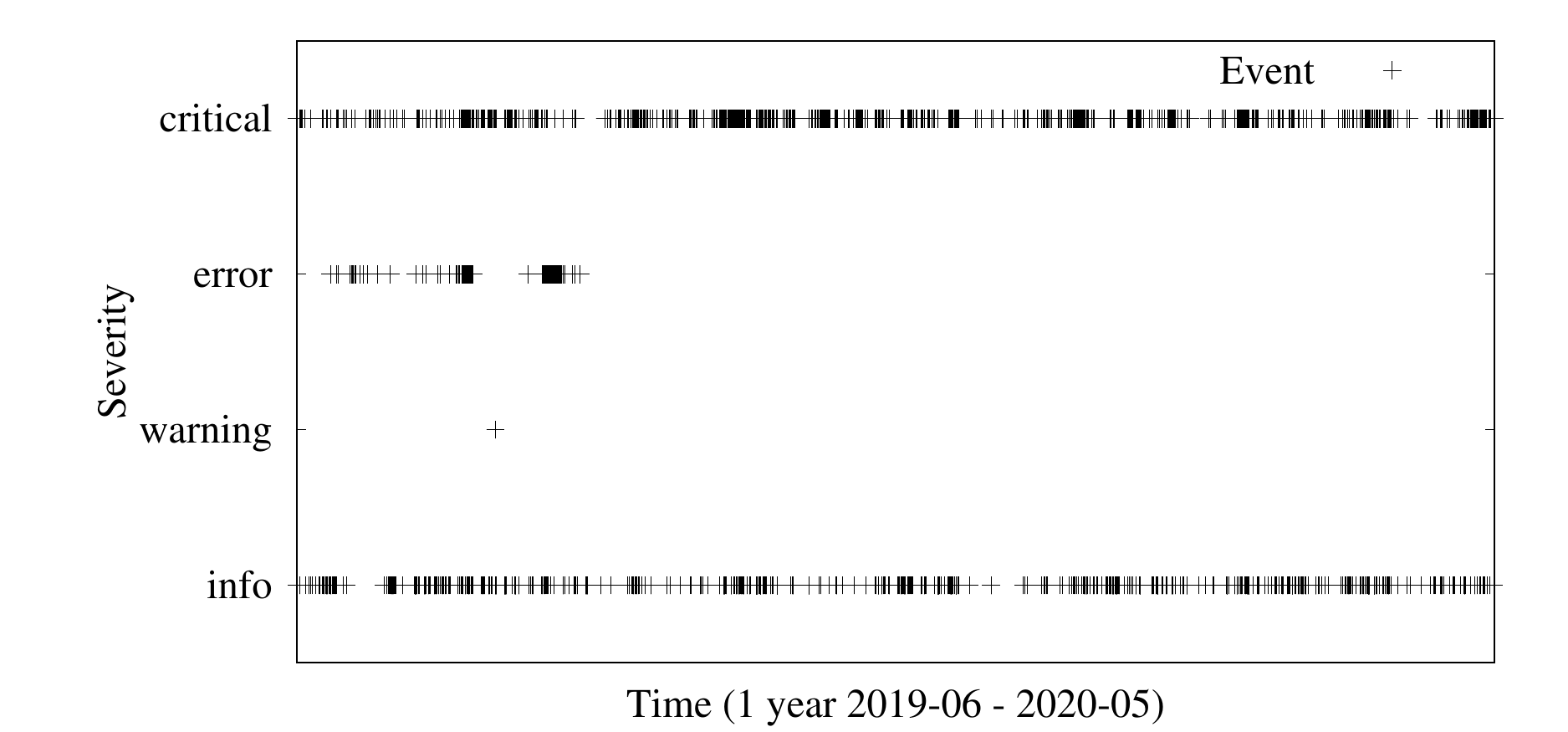}
\caption{Alerts over 1 year with severities for a production application}
\vspace{-3mm}
\label{fig:events_one_year}
\end{figure}

As an example, Figure~\ref{fig:events_one_year} shows 1 year worth of alerts
(only escalated events) generated from one production application that is
composed of 26 microservices running on Kubernetes clusters.  The alerts
include 10,399 (75.79\%) critical, 367 (2.68\%) error, 3 (0.02\%) warning, and
2,951 (21.51\%) info. alerts.

\subsection{Incidents / Change Records}

A manually or automatically created {\bf incident} serves as a formal record
to log all the information relevant to an issue and resolution.  An
incident record typically captures the information like: id, title,
description, opened date, severity, impacted configuration item(s), outage
start time, outage end time, state (resolved, open), resolution
description, change ID to represent the incident has been ``caused by
change''.  A {\bf change record} captures key attributes for a change like:
id, title, description, purpose, environment, request date, start date,
end date, team, state (open, closed), closure code
(successful, failed, induced issues), backout plan, close
notes, configuration items for resources or images associated with the change
ticket.  Incidents and change records are often used to predictively measure
the potential risk of the submitted change (not yet deployed) or reactively
find which images have been problematic.

Any changes---for instance, performed to add new features, address
vulnerabilities or improve the performance of the system---can also induce
alerts/incidents. Any changes to the software are generally made by modifying
the source code, rebuilding the images and redeploying the newer version of the
images to containers (\S\ref{subsec:topology}).  So, when an alert or group of
alerts is produced on any component of the virtualized layer, checking for
recent changes deployed on the component can help in the root cause analysis.

Observer tools to monitor the CICD pipeline can help link changes directly to the
DevOps workflow, but in absence of such a mechanism, we will have to extract
references to image names from the change tickets. While in some cases, information
about the image(s) deployed by change can be mentioned in structured fields of the
change ticket, in most cases the reference to images is hidden in the unstructured
text fields. We map change tickets to topology objects by mining the references to
image names from the change text. We first look for direct mentions of image names
in the change tickets. If direct mentions are not identified then we search for
reference to image tags as it is unlikely for two images to have the same tag.
If both of these return no matches, then we search for similar changes to the
current change ticket and add the image reference if the similar changes have
any images mapped to them.

Based on the above image mapping, we identify the virtual topology objects
associated with each change ticket and add change ticket references to the
alert(s) mapped to the same topology object through entity resolution
(\S\ref{subsec:entity_resolution}) if the change was recently deployed before
the alert was generated.

\begin{table}[t]
\scriptsize
\begin{center}
  \begin{tabular}{ | m{0.6cm} | m{1.2cm} | m{5.8cm} |}
    \hline
    {\bf Type} & {\bf Category} & {\bf Relationships} \\ \hline \hline
    Node & Type & application, backplane, bridge, card, chassis, command, component, container, cpu, database, directory, disk, emailaddress, event, fan, file, firewall, fqdn, group, host, hsrp, hub, ipaddress, loadbalancer, location, networkaddress, networkinterface, operatingsystem, organization, path, person, process, product, psu, router, rsm, sector, server, service, serviceaccesspoint, slackchannel, snmpsystem, status, storage, subnet, switch, tcpudpport, variable, vlan, volume, vpn, vfr \\ \hline
    Edge & Aggregation & contains, federates, members \\ \hline
    Edge & Association & aliasOf, assignedTo, attachedTo, classifies, configures, deployedTo, exposes, has, implements, locatedAt, manages, monitors, movedTo, origin, owns, rates, resolvesTo, realizes, segregates, uses \\ \hline
    Edge & Data flow & accessedVia, bindsTo, communicatesWith, connectedTo, downlinkTo, reachableVia, receives, routes, routesVia, loadBalances, resolved, resolves, sends, traverses, uplinkTo \\ \hline
Edge & Dependency & dependsOn, runsOn  \\ \hline
    Edge & metaData & metadataFor \\ \hline
  \end{tabular}
   \caption{Entity and edge semantic relationships in topology service}
   \vspace{-4mm}
    \label{table:topology_types}
   \end{center}
\end{table}

\subsection{Virtual Topology}
\label{subsec:topology}

In legacy systems, a (physical) node is often equivalent to an application (or
multiple processes running in the same node) or one database and physical
interfaces thereof represent connectivities with other nodes (applications).
In today's cloud systems, a physical node is virtualized to run multiple
\emph{virtual machines (VMs)}, even further each VM is virtualized to run
multiple \emph{containers}.  Also, the virtualized network provides virtual
interfaces that support overlay or hybrid protocols.
Figure~\ref{fig:virtualization} illustrates evolution from legacy systems
(left) to contemporary cloud systems (right).

\begin{figure}[t]
\center
\includegraphics[bb=0 0 100 100, trim=0.45cm 0 0 6.8cm, clip, width=0.4\textwidth]{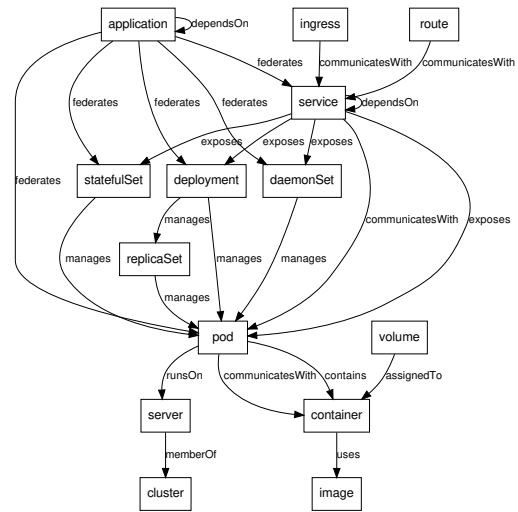}
\vspace{-6mm}
\caption{Topological entities and relationships from cloud deployment}
\vspace{-3mm}
\label{fig:topology}
\end{figure}

To correctly diagnose a problem, identifying where alerts are generated is the
key to understand the problem correctly.  The systems are often represented as
\emph{topological graphs} that have nodes (i.e., any box in
Figure~\ref{fig:virtualization}), and edges (i.e., vertical connectivity or
horizontal connectivity).  Table~\ref{table:topology_types} shows types of
nodes and edges representing cloud systems: 52 node types and 41 edge types.
The semantic relationships among nodes and edges are quite complex as shown in
Figure~\ref{fig:topology}. Therefore, identifying `correct' topological
entities (\S \ref{subsec:entity_resolution}) and correlating alerts derived
from them (\S \ref{subsec:correlation}) are the main stepping stones for SREs
to tackle the problem.


\subsection{Enrichment}
\label{subsec:enrichment}
Artifacts like alerts or change tickets generally make reference to topological
entities to which they relate.  Knowing which topological entities are involved
is necessary for problem localization, and is a general assistance to any
correlation effort.  These entities are not always explicitly listed in a
well-formatted way, although they sometimes are.  They may be included in a
JSON object that has been converted to a string in an embedded field that is
standard in a particular deployment, referenced by name in a field provided by
a performance monitor template, or identified by IP address in the free text of
the description.

\cut{ 
For this reason, we use a configuration file that looks at commonly used
locations, and allow each deployment to extend it with fields used in that
environment.  When a topological reference has been extracted, it is validated
by searching the topology service, which returns a unique identifier.  This
simultaneously avoids false references and unifies different formats like
address and DNS name.
}

Logs and metrics are two key sources of information that an application
generates, irrespective of whether it is in healthy state or unhealthy state.
Moreover, the logs generated by an application are in a finite space such that
they can be mined and mapped to a set of template-ids. In order to use logs for
event correlations, each log line is processed and templatized using a pipeline
called as {\it log-template pipeline}. The log-template pipeline consists of
two components, error classification of log line followed by templatization of
log line.  When any event is generated, its corresponding log lines that are
within a fixed duration of the start of the event are fetched. Each log line
from the set of log lines is input to a pre-trained classifier, the output of
the classifier is a 0 (error) or 1 (non-erroneous). The error classifier allows us to
separate log lines pertaining to healthy state of the system and the
corresponding microservice from the non-erroneous log lines. The outcome of the
error classifier is a subset of total log lines which are then input to the
next step of the pipeline, {\it template miner}. A template miner is pre-trained on
millions of log lines that can map a log line to a template id.  For each
erroneous log line, we obtain its corresponding template-id from the template
miner, thus yielding a set of template-ids for all log lines. Each event is
enriched with its entities and a set of template-ids. The entities and
template-ids contained in the enriched event are used at a later stage for
event correlation.
\subsection{Normalization}

Various data sources (events) flowing into \name are normalized or standardized
with the same format in order to increase the cohesion of data types and reduce
the redundancy. Having looked through many data sources, we have found that the
following information is enough to run all the algorithms: title, description,
created\_at, resolved\_at, severity, source, and features. Note that features
are expandable to accommodate any unique information from any data source as  
an object such as name, URL, alert\_id, team, and application.

\section{Methodology}
\label{sec:methdology}

\subsection{Hybrid Entity Resolution}
\label{subsec:entity_resolution}

\name uses multiple sources for operational events.  Some are
more easily associated with a specific topological entity than others.
For example, if our log anomaly detection subsystem creates an event to alert
operations of unusual behavior, then it can label that event with the exact
container that generated the log being observed.  If, on the other hand, we
import an alert from an external alert management system, which in turn created
it in response to a level-crossing observed by a monitoring tool, matching it
with the topology is harder.

The first challenge is that these alerts are often created manually, and how
they identify the resource depends on local standards.  To allow the
embedded information to be used for locating the resources, we have defined a
``matcher'' language ({\it template}) for describing how these locations may be
embedded in the raw alert object.  Alert management systems have some standard
fields, so our starting point is a standard extraction file that checks these
known places.  Customers who follow local practices, such as a standard
name-value pair in the title of the alert, can add this to the
template rules file.

In addition to this template-based approach, we also use a dictionary-based
approach to scan for things that ``look like'' entity references.
Specifically, we know the domain name server (DNS) in use in the subject
environment.  We also know a number of names used by docker images and
Kubernetes objects in this environment.  These are all partial names, not
knowing every host name or every image tag, but by combining the names that we
do know with regular expressions for how they generally appear allows us to
search for likely references.  To adapt to the changing environment, we use an
internal query system that allows us to compile a rapid regular expression
matcher for fast application, while still allowing the specific list of names
({\it dictionary}) to be updated dynamically.

The second challenge is that resources may be identified using different
methods, and no single method will work for all resources.  In our case, we use
resource IDs (e.g., UUID), assigned by a topology service when the topological
object is added to its database, but an external monitoring tool will not know
these IDs.  Kubernetes has unique object IDs, but they are also generally
invisible to an external monitor.  IP addresses and DNS names identify an
endpoint, but that is often just a front end, such as a Kubernetes ingress or a
load balancer.

Once we have used the rules file to get a value that should identify an object
in the topology, we then use a search to resolve that reference to an object
ID.  Topology searches need to identify the time at which to search, so we use
the creation timestamp on the alert.  If the alert was actually a result of the
object being deleted, then it did not exist at the creation time of the alert,
so we will not resolve that object.  In practice, this does not appear to be a
significant problem.

\subsection{Learning Patterns}
\label{subsec:learning_patterns}

The same types of problems tend to happen repeatedly and capturing them as
patterns help the problem determination.  In this section, we introduce various
ways to capture patterns for the given datasets.


\noindent {\bf Association rule mining}: one of the most important data mining
algorithms, aims to discover interesting relationships latent in large
datasets~\cite{han2011data}. A typical and widely known example of association
rules application is market basket analysis by learning from sales
transactions. The strategy of association rule mining is composed of two
phases~\cite{tan2016introduction}: 

\begin{noitemize}
\item{\textbf{Frequent Itemset Generation}: The objective of this phrase is to find all the itemsets, also called frequent itemsets, if items occurred together greater than a minimum support threshold ($min\_sup$) in transactions.} 
\item{\textbf{Rule Generation}: This phrase is to extract all the high-confidence rules (i.e., strong rules) based on the minimum confidence threshold ($min\_conf$)) the frequent itemsets generated from the first phrase.} 
\end{noitemize}

\begin{figure}[!ht]
\center
\includegraphics[bb=0 0 100 100, trim=0 0 0 0, clip, width=0.5\textwidth]{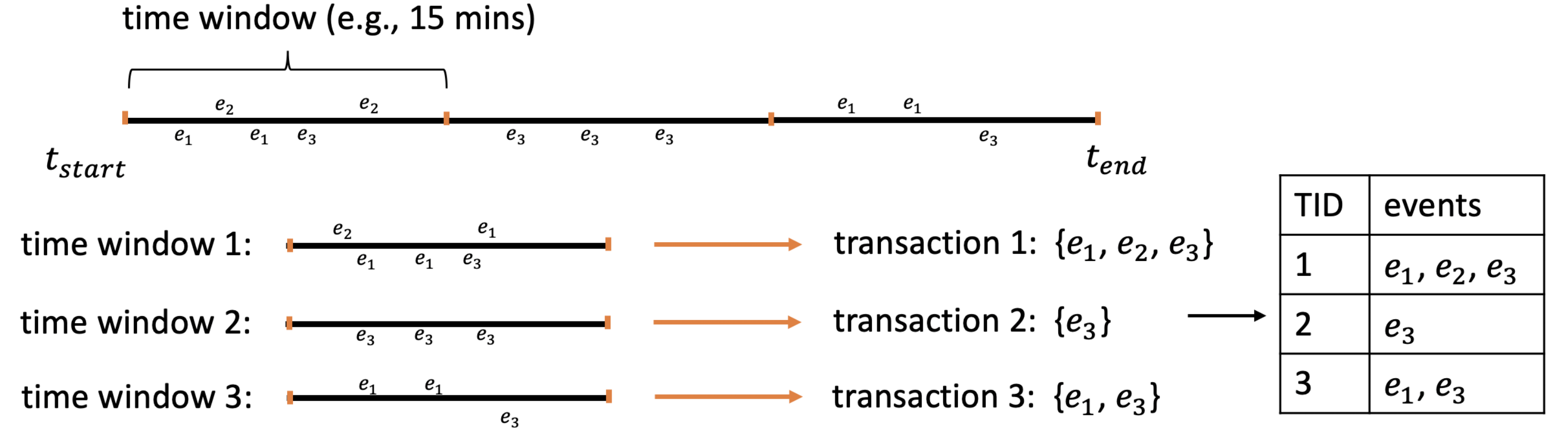}
\caption{An example of converting sequential events to \textit{events transactions}. In this example, three types of events (i.e., $e_1$, $e_2$, and $e_3$) are monitored and reported from time $t_{start}$ to time $t_{end}$.}
\label{fig:events-to-transactions}
\end{figure}

A fixed time window (e.g., 10 mins) is applied to generate \textit{events transactions} from the sequential events monitored from $t_{start}$ to
$t_{end}$ (seen in Figure~\ref{fig:events-to-transactions}). Each row in the
table corresponds to a transaction containing a unique identifier labeled TID
and a set of events in a time window. In this example, let $E=\{e_1, e_2,
e_3\}$ be the set of all event types and $T=\{t_1, t_2, t_3\}$ be the set of
all transactions. Supposing $min\_sup=50\%$ and $min\_conf=80\%$, we could get the
frequent itemsets $\{e_1, e_3\}$ and strong rule $\{e_1 \rightarrow e_3\}$
through the calculations shown as follows:

\vspace{-4mm}
{\footnotesize
$$
\text{Support}(\{e_1,e_3\}) = \frac{\sigma(\{e_1, e_3\})}{|T|} > min\_sup, 
$$
where $\sigma(X) = \{t_i|X \subseteq t_i, t_i \in T \}$, and
}
 
{\footnotesize
$$
\text{Confidence}(e_1 \rightarrow e_3) = \frac{Support(e_1, e_3)}{Support(e_1)} > min\_conf. 
$$ 
}
\vspace{-4mm}

The uncovered correlations (i.e., frequent itemsets and strong rules) can perform basis for decision making and prediction to support hybrid cloud management such as event correlation, anomaly detection, fault localization, and the like.

\noindent {\bf Log templates}:
Two events may or may not have a similar description, however, if the
underlying logs are similar then they are most likely related to each
other---this is the key hypothesis of using logs for event correlation. 

Each application consists of several microservices, some of these services are
related to other services forming a graph. If one service fails then any other
service which is upstream or downstream of the failed service could throw
error log lines. It is important to identify error log lines for each failed
microservice during an execution of an application, collate them together to
form log signature for a particular event. 

In order to use logs for event correlations, each log line is processed and
templatized, then they are collated to form a log-signature for each event. To
recall, from \S\ref{subsec:enrichment} the log-template pipeline consists of
two components, error classification of log line and templatization of log
line. As a result, for a given event, there is a set of templates-ids and
corresponding application-ids. We propose a log-signature representation for
each event from its template-ids and corresponding application-ids, and use
that for event correlation. The example below shows a log signature for an
event; there are three log template ids---\texttt{template\_id\_a},
\texttt{template\_id\_b}, \texttt{template\_id\_c}; two log template ids
(\texttt{template\_id\_a} and \texttt{template\_id\_b}) belong to
\texttt{application\_id\_a}, and one log template id (\texttt{template\_id\_c})
belongs to \texttt{application\_id\_b}. This representation is called as log
signature of an event. 

\vspace{-2mm}
\begin{verbnobox}[\fontsize{7pt}{7pt}\selectfont]
{
  "templates": [{
      "application_id": "application_id_a",
      "template": "template_id_a"
  }, {
      "application_id": "application_id_b",
      "template": "template_id_b"
  }]
}
\end{verbnobox}
\vspace{-2mm}

Once we have a log signature for each event, 
the similarity is calculated between two events by computing the overlap
between their application ids; for each application id that overlaps, it
computes the overlap between their respective templates ids to calculate a
score called as log template similarity score. For event grouping using
template clustering, the algorithm starts with a list of groups, each group
could contain one or more events. Next, it computes the similarity between each
pair of groups using the log template similarity, as explained above. This
produces pairs of groups with similarity score between them, the pair with the
highest similarity score which is above a threshold is taken and
the groups in the pair are merged into one single group.  Then the process is
repeated until there is only one group left or the highest similarity score
between groups is below the threshold.


\begin{figure}[t]
    \centering
    \begin{minipage}[b]{0.45\linewidth}
        \includegraphics[bb=0 0 100 100, trim=0 0 0 0, clip, width=1.1\textwidth]{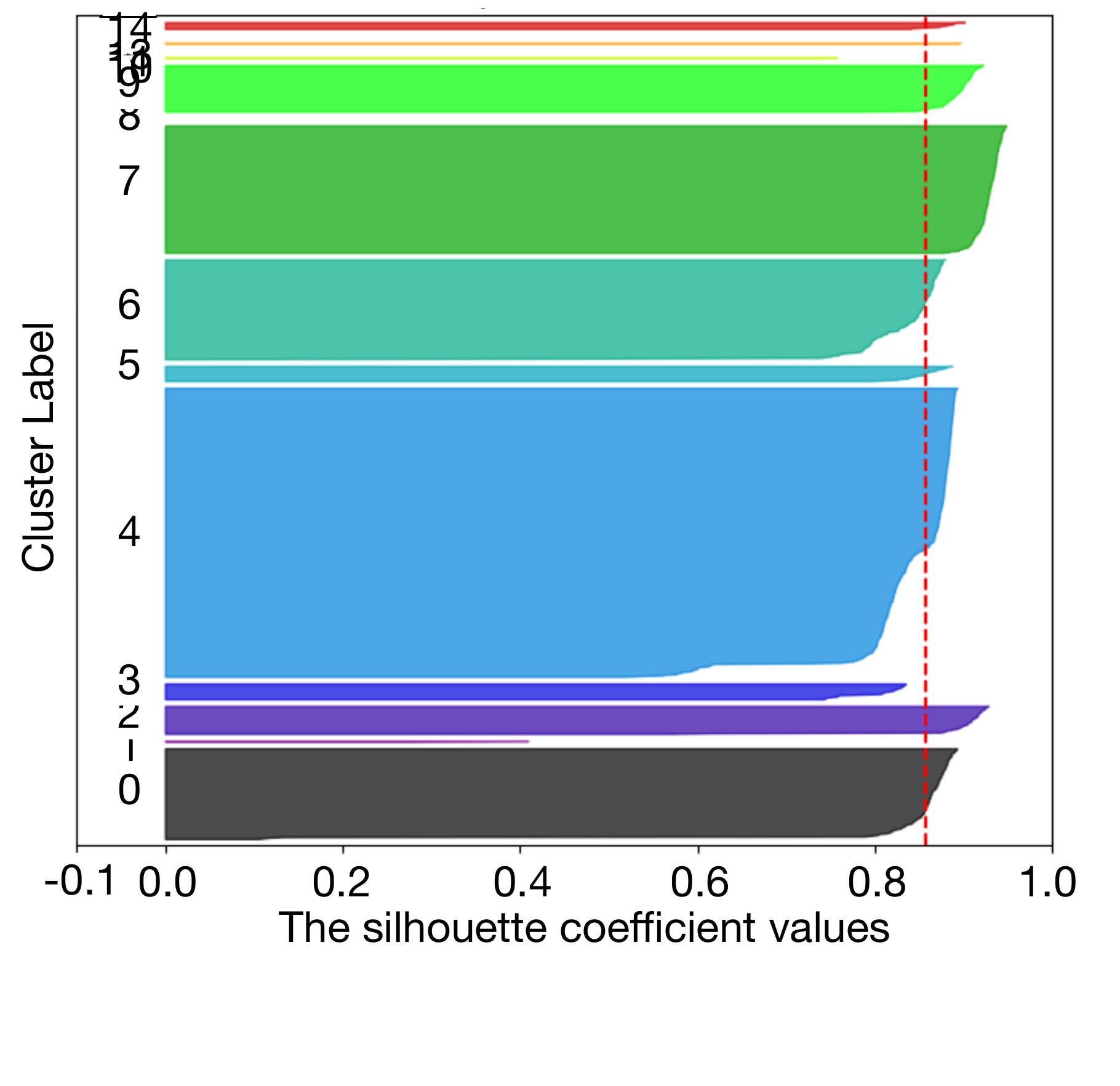}
        \vspace{-10mm}
        \caption{Silhouette coefficients in case of 15 clusters}
        \label{fig:silhouette}
    \end{minipage}
    \quad
    \begin{minipage}[b]{0.45\linewidth}
        \includegraphics[bb=0 0 100 100, trim=0 0 0 0, clip, width=1.1\textwidth]{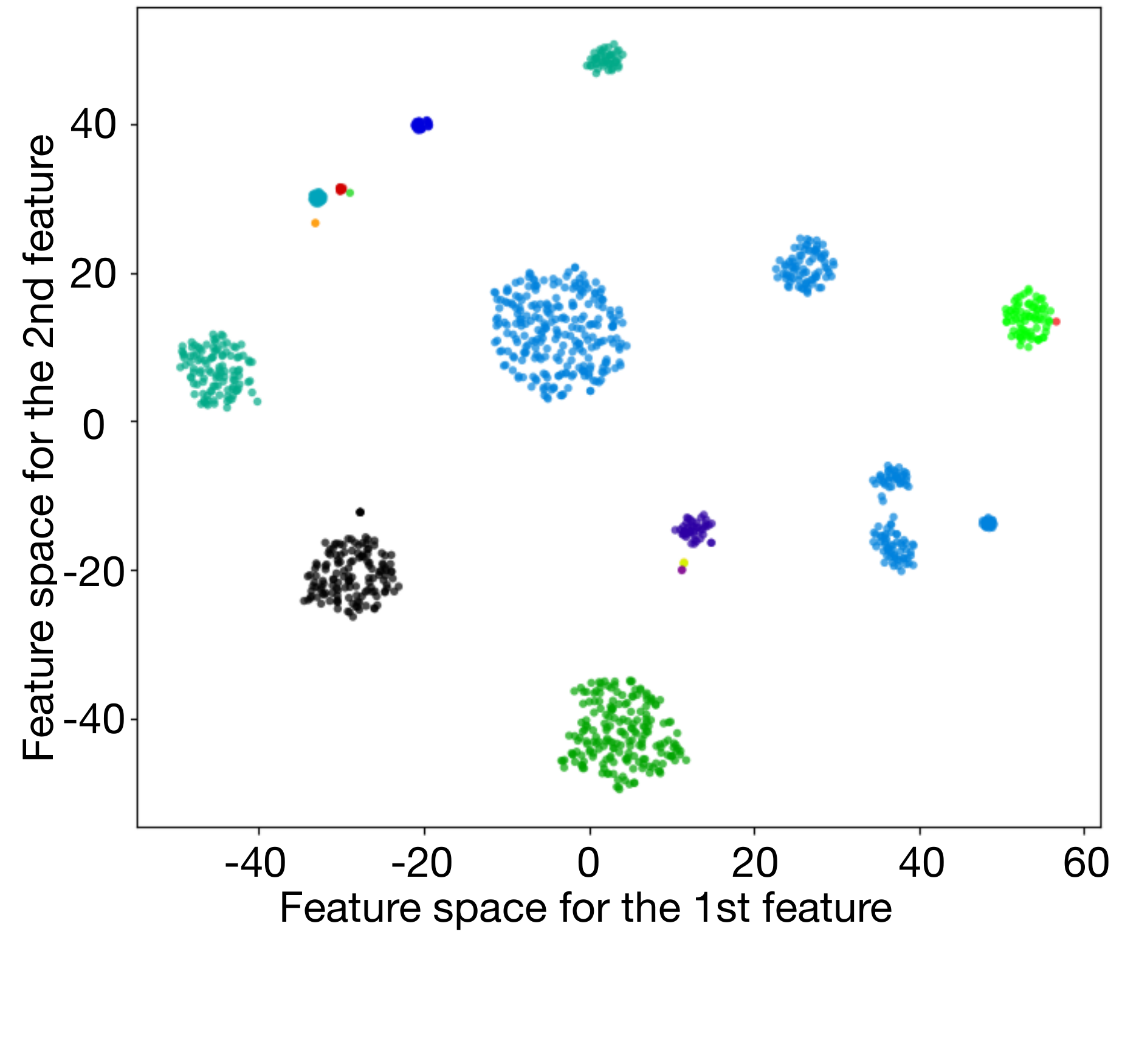}
        \vspace{-10mm}
        \caption{t-SNE for word embeddings on alerts and 15 clusters}
        \label{fig:tsne}
    \end{minipage}
    \vspace{-4mm}
\end{figure}

\cut{
\begin{figure}[t]
\center
\includegraphics[bb=0 0 100 100, trim=120 0 650 60, clip, width=0.4\textwidth]{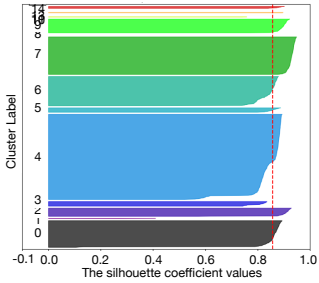}
\caption{Silhouette coefficients in case of 15 clusters}
\label{fig:silhouette}
\end{figure}

\begin{figure}[t]
\center
\includegraphics[bb=0 0 100 100, trim=650 0 120 60, clip, width=0.4\textwidth]{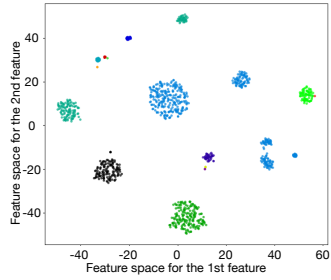}
\caption{t-SNE for word embeddings on alerts and 15 clusters}
\label{fig:tsne}
\end{figure}
}

\noindent {\bf Word embeddings}: 
Application events, different from network events, are not often structured in
that the data descriptions or log messages are written in natural language. So,
the natural language processing would help understand the patterns of the
events.  Because learning word embeddings is effective to capture the same or
similar representation for words that have the same
meaning~\cite{mikolov2013efficient}, the
fastText (\url{https://fasttext.cc}) word embedding (vector size = 300) has
been  applied to the alert descriptions to train the embedding model. Later,
this model turns alert descriptions into learned representations.  In order to
identify how the embedding is clustered, a balanced iterative reducing
and clustering using hierarchies (BIRCH) is applied to the
embeddings~\cite{10.1145/235968.233324}. 

The silhouette coefficient is a measure $[-1, 1]$ of how close each point in
one cluster to points in the neighboring clusters and thus provides a way to
assess goodness of clusters. Figure~\ref{fig:silhouette} shows the silhouette
coefficients. The average silhouette coefficient is the maximum when the number
of clusters is 15. This means the quality of clustering is the best when 15
clusters are formed. This will be further evaluated in the evaluation
(\S\ref{subsec:eval_feedback}).  Also, for the same number of clusters,
Figure~\ref{fig:tsne} shows the clustering space in 2 dimensions with
\textit{t-SNE} that is a technique of non-linear dimensionality reduction and
visualization of multi-dimensional data.

\begin{table}[h]
\scriptsize
\begin{center}
  \begin{tabular}{| m{3.5cm} | c | c | c |}
    \hline
    {\bf -} & {\bf Similarity} & {\bf Knowledge} & {\bf Statistical} \\ \hline \hline
    Group alerts from diff. sources & O & O & X \\ \hline
    Require prior knowledge & O & O & X \\ \hline
    Detect false alerts & O & O & Guess \\ \hline
    Detect multi-stage incidents & X & O & Guess \\ \hline
    Find new incidents & O & X & O \\ \hline
    Error rate & Mid & Low & High \\ \hline
  \end{tabular}
   \caption{Qualitative comparison for different approaches}
   \vspace{-4mm}
   \label{table:correlation_comparison}
   \end{center}
\end{table}

\subsection{Hybrid Correlation}
\label{subsec:correlation}

As briefly discussed in \S \ref{sec:introduction} and \S \ref{sec:related},
generally similarity (e.g., rules, patterns), knowledge (e.g.,
scenario, knowledge base), and statistical (e.g., statistical estimation,
causal relation estimation, reliability degree combination) approaches are used
for the information
correlation~\cite{Mirheidari_2013,10.1109/COMST.2016.2570599}.
Table~\ref{table:correlation_comparison} compares the three approaches
qualitatively. Not one approach is perfect, but an hybrid approach would
complement deficiencies.

\begin{figure}[t]
\center
\includegraphics[bb=0 0 100 100, trim=0 0 0 0, clip, width=0.5\textwidth]{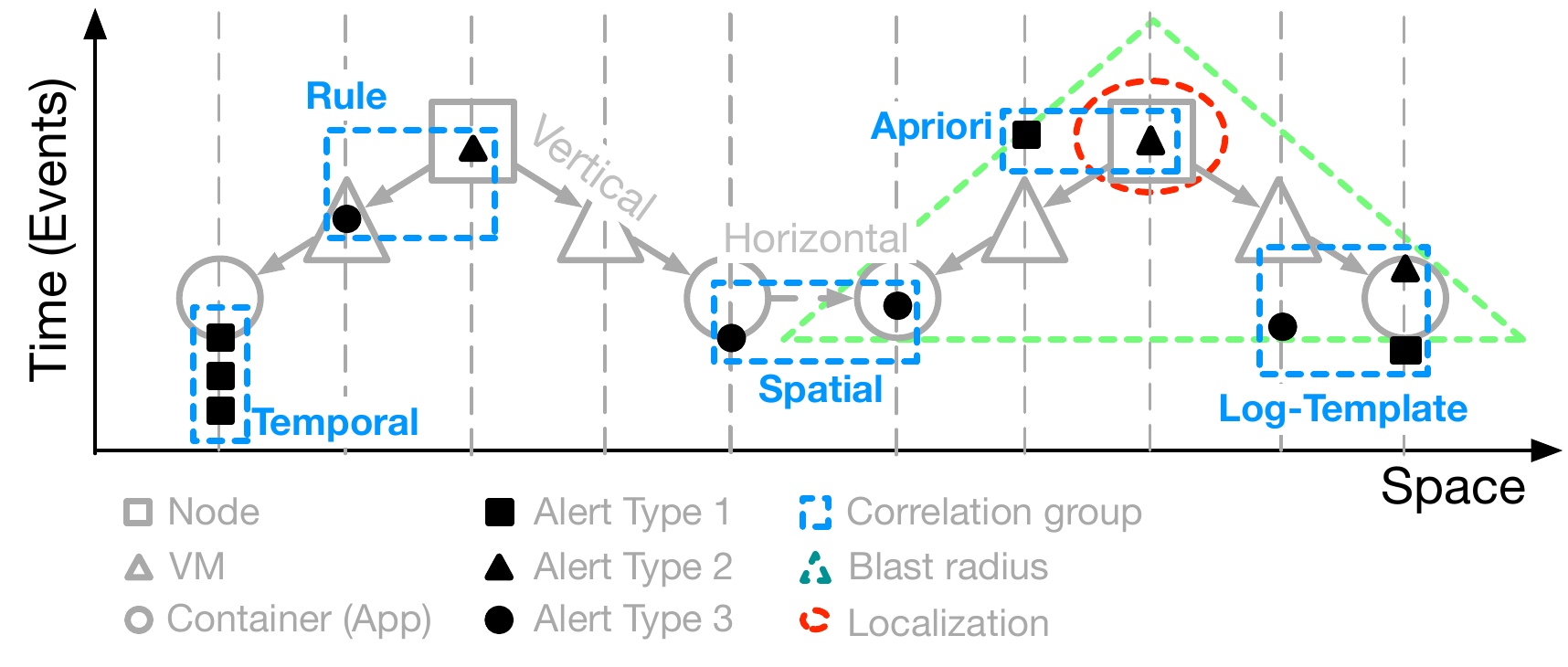}
\caption{Illustration of correlated alert groups with different methods}
\label{fig:correlation}
\end{figure}

\name uses a mix of different methods to make a correct verdict on the
correlated information.  For \emph{similarity}, rules (time, spatio, prior
information), patterns/templates (\S \ref{subsec:learning_patterns}), and for
\emph{knowledge}, knowledge base (capture causal inference from SREs or root
cause analysis), feedback (\S \ref{subsec:continuous_improvement}), and lastly
for \emph{statistical}, machine learning algorithms (e.g., association rule
mining), clustering (\S \ref{subsec:learning_patterns}) are used.
Figure~\ref{fig:correlation} illustrates some of methods in time (y-axis) and
spatial (x-axis) dimensions.  \Square, \TriangleUp, and \Circle \; represent
spatial information (\S\ref{subsec:topology}), respectively, physical nodes, VMs
and containers that have vertical relationship with ``runsOn'', and horizontal
relationship with ``dependsOn'' (Table~\ref{table:topology_types}).
\FilledSquare, \FilledTriangleUp, and \FilledCircle \; represent different
types of alerts.

\textcolor{blue}{dotted-\Square} describes correlated alert groups that are
made by different methods.  The \emph{temporal} group is based on the time
given that alerts are generated from the same topological entity. Similarly,
the \emph{spatial} considers both time and space where entities have a special
semantic relationship (in this case ``dependsOn'').  The \emph{rule} group is
derived from SREs' input, \emph{apriori} group is based on the patterns learned
from the association rule mining, and the similarity of \emph{log-template}
group is measured by templates of alerts (logs). We demonstrate the
effectiveness of this ensemble approach in \S
\ref{subsec:eval_event_correlation}.

\subsection{Localization}
\label{subsec:explainability}

The correlated alert group itself explains about the problem, but often case,
it is not enough for software developers and SREs to understand the root cause
immediately, especially when the group includes multiple alerts from different
locations.  Additional explainability would help localize the problem and
understand the impact of it.  Treating alerts in the group as error signals and
knowing the topological entities and their relationships render reasoning to
find the root entity of the problem and its impact for the related entities.
In Figure~\ref{fig:correlation}, \textcolor{red}{dotted-\Circle} shows the
localized entity given the correlated group based on the apriori.
\textcolor{green}{dotted-\TriangleUp} is a blast radius (impact) of the
problem.  While SREs know well about the architecture of their application, the
localization and blast radius help SREs to visualize the problem together with
the description of the alerts.  \name uses a
Souffle\footnote{https://github.com/souffle-lang/souffle} reasoner to traverse
topological entities with error signals and outputs topological entities for
localization and blast radius.


\subsection{Continuous Improvement (Feedback)}
\label{subsec:continuous_improvement}

The feedback is an essential part of continuous learning for AIOps systems, but
collecting feedbacks is challenging with the following reasons: first,
collecting a large number of feecbacks is hard as SREs are not committed or
incentivized; second, correctness or consistency is not guaranteed due to
ambiguity from different experiences or skillsets; third, details are often
missing as the most effective way is yes/no or thumbs up/down or at best
drop-down with a short list of selections. Especially, since the ChatOps
interface (main interface for \name) flows based on timeline rather than
dashboard, it is harder to wait SREs to finish the work and come back to
provide feedbacks.  Therefore, learning from minimal feedbacks (i.e., active
learning) is important.  The feedback is expected to improve the correctness of
correlation and suppress false positives.  \name uses a content-based approach
that leverages clustering \emph{split} and \emph{merge} operations based on the
word embedding~\cite{balcan_blum_2008} (\S\ref{subsec:learning_patterns}).
Applying feedbacks improves correctness of rules, patterns, and knowledge base. 

\cut{
Content-Based filtering (CBF)
Collaborative filtering (CF)
User-Based Collaborative Filtering (UBCF)
Item-Based Collaborative Filtering (IBCF)

Operations to apply feedback for unsupervised:
Split: divide into two groups
Merge: combine two groups into one group
Join: add one or more event(s) to a group
Separate: exclude one or more event(s) from a group

Online algorithms decide event groupings with feedbacks reflected in the next decision for the same event groups
Offline algorithms take feedback as inputs for retraining and get better (i.e. more accurate) results
}


\section{Evaluation}
\label{sec:evaluation}

\name enables hybrid diagnosis approaches on various data types to correctly
identify entities and group information for software engineers and SREs to
diagnose problem without spending too much time for post-mortem analysis.  In
this section, we evaluate \name with the following goals:

\begin{noitemize}
\item Demonstrate a hybrid approach for data enrichment with entity resolution and show the effectiveness (\S \ref{subsec:eval_data_enrichment}),
\item Show how hybrid correlation approaches improve correctness (\S \ref{subsec:eval_event_correlation}), and
\item Analyze how minimal feedbacks can improve the correlation results (\S \ref{subsec:eval_feedback}).
\end{noitemize}


\subsection{Setup}
\label{subsec:setup}

A total of 12,696 alerts (1 year) have been obtained from the product team that
manages a software as a service (SaaS) application in IBM Cloud\footnote{The application name and data are considered IBM confidential, so it is not included in this paper}.  Our algorithms have been applied to 1,134 alerts for a worth
of 1 month and identified 382 issues (groups)---141 issues have more than 1 alert and
the rest 241 includes only single alert.  This dataset has been shared with
developers and SREs to get feedbacks for each alert whether it belongs to the
correct group or not.  The rest of 11,565 alerts have been used for training to
learn patterns.  In addition, the snapshot of the virtual topology running on
Kubernetes has been shared.  It includes 516 nodes, 755 deployments, 1,532 pods
and 801 services.

\subsection{Hybrid Entity Resolution}
\label{subsec:eval_data_enrichment}

Our two main extraction techniques, described in
\S\ref{subsec:entity_resolution}, are template-based and dictionary-based
approaches.  As the labeled dataset does not include the ground-truth of the
topological entities, we use the sampling and manual labeling.  When deciding
on which to incorporate into the final system, we start with a random sample of
10 events from the dataset, and have manually identified 16 useful topological
references that they contain (true positives).  The template-based approach
finds 9, and the dictionary-based approach finds 10, so the combined approach
returns 15 correct (93\% accuracy).  However, only 4 are found by both.
Although it is a small set, the results gave us a clear indication that neither
technique alone would likely be sufficient.

\subsection{Hybrid Correlation}
\label{subsec:eval_event_correlation}


As described in \S\ref{subsec:correlation}, the hybrid correlation would
benefit toward correctness. In this section, we show how applying various
methods help improve performance.

\noindent {\bf Temporal and spatial}:
Without learned models that help identify similarity, knowledge, statistical patterns,
the time and space are base attributes towards the correlation. Out of 1,134
alerts (forming 382 groups), based on time and space, 825 alerts labeled as correct
(TP), 68 as incorrect (FP), 241 (single alert in group) as not applicable (FP).
Since the data is not necessarily associated problems, feedbacks about what
additional data should be included (FN) have not been collected\footnote{True
Positive (TP), False Positive (FP), False Negative (FN)}.  Therefore, 

\vspace{-2mm}
{\footnotesize
$$ 
Precision (P) = \frac{TP}{TP + FP} = \frac{825}{1134} = 0.73
$$
$$
Recall (R) = \frac{TP}{TP + FN} = \frac{825}{825} = 1.00
$$
$$
F1Score = \frac{2 \times R \times P}{R + P} = \frac{2 \times 1.00 \times 0.73}{1.00 + 0.73} = \frac{1.46}{1.73} = 0.84
$$
} \vspace{-4mm}

Given the labeled data without knowing true/false
negatives, precision is important to understand correctness.
This is our baseline performance.

\begin{figure}[t]
\vspace{-4mm}
\centering
    \begin{minipage}[b]{0.45\linewidth}
        \includegraphics[bb=0 0 100 100, trim=0 0 0 0, clip, width=1.2\textwidth]{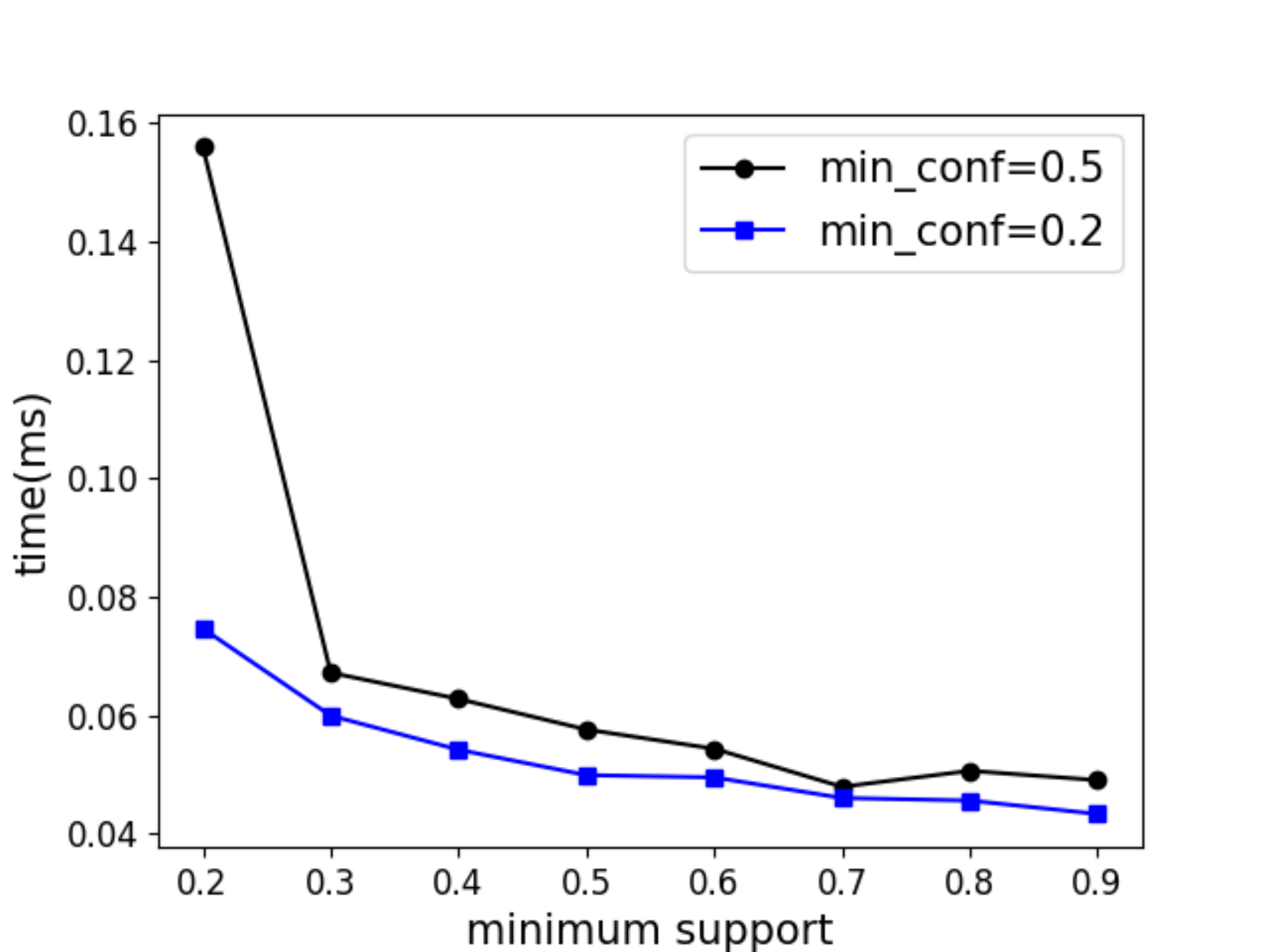}
        \vspace{-5mm}
        \caption{Running time for generating frequent itemsets and associations rules at different minimum support levels}
        \label{fig:min_support_tc}
    \end{minipage}
    \quad
    \begin{minipage}[b]{0.45\linewidth}
        \includegraphics[bb=0 0 100 100, trim=0 0 0 0, clip, width=1.2\textwidth]{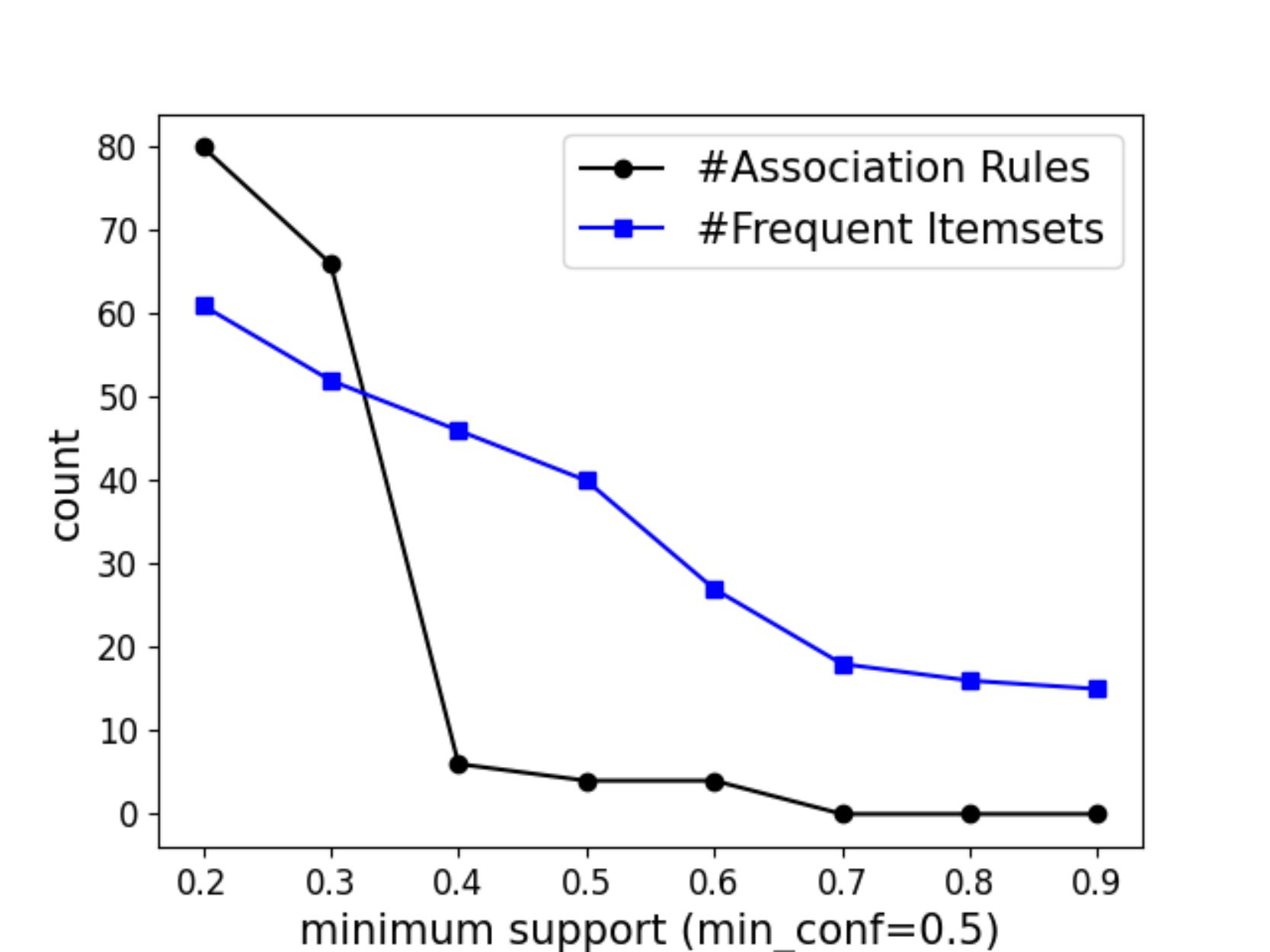}
        \vspace{-5mm}
        \caption{The number of frequent itemsets and associations rules at different minimum support levels with $min\_conf=0.5$}
        \label{fig:min_support_rules_items}
    \end{minipage}
    \vspace{-4mm}
\end{figure}

\noindent {\bf  Association rule mining}:
In this section, we explain how we apply {\it apriori}, one of the most popluar
association rule mining algorithms, to find frequent co-occurrence alerts
(i.e., events) from our dataset, which are generated from 73 sources (i.e.,
pods, containers, nodes). Our goal is to find associations among these alerts
from each source. As detailed in \S\ref{subsec:learning_patterns}, we first get
alert transactions from sequence alerts and use apriori to generate frequent
itemsets and rules for each source. Since each source is not large enough, 5
mins is used as the fixed time window and finally we get 852 transactions in
total. The maximum number of transactions in sources is 52 and the average
numer is 12. The performance curve of running time for generating frequent
itemsets with minimum supports from 0.2 to 0.9 is shown in
Figure~\ref{fig:min_support_tc}. Figure~\ref{fig:min_support_rules_items}
indicates that the number of frequent itemsets and the number of association
rules increases as the minimum support is reduced on this data. In order to
reveal how apriori can improve the correlation performance, we select a minimum
support threshold $0.3$ to generate frequent itemsets. When $min\_sup=0.3$,
there are 4 $L_2$ and 2 $L_3$ frequent itemsets, where $L_k$ indicats the
length of itemsets is $k$, among all generated 52 frequent itemsets. Based on
the feedback from SREs, all the learned frequent itemsets are the correct
groups. Using these frequent itemsets (i.e., clusters), we split a group into
multiple smaller groups or merge multiple clusters into a larger one in order
to improve the accuracy of groups. As expected, 31 alerts, out of 68 incorrect
labeled alerts from temporal and spacial group results, are split and merged
into another groups or become a new correct group. the precision of correlation
improves to $0.78$, while F1 improves by $3.5\%$.

\noindent {\bf  Log-template grouping}:
In this section, we want to verify the hypothesis that the two similar alerts may or
may not have a lexically matching incident descriptions, but their logs should
have high overlap and that they are discriminative. In order to validate the
hypothesis, a total of 10 similar pairs and 19 dissimilar pairs are sampled
from the dataset with their description, log-templates and entities. For
each pair of alerts, we compute the similarity score between them based on:
alert-description, entity, log-templates. To compute the alert-description
based similarity between two alerts, we obtain the distributed representation
using universal sentence encoder for each alert in the pair, and then compute
cosine similarity between them. For entity based similarity, we first extract
entities from each alert, represent them as a vector of entities. Then, we
compute the tf-idf similarity between them. \S \ref{subsec:learning_patterns}
outlines our method for calculating log-template based similarity between two
alerts.

\begin{table}
    \centering
    \scriptsize
    \begin{tabular}{|l|l|l|l|l|l|l|}
    \hline
         & \multicolumn{2}{c|}{Log-template} & \multicolumn{2}{c|}{Entity} & \multicolumn{2}{c|}{Event-desc} \\ \hline
         & Similar & Dissim. & Similar & Dissim. & Similar & Dissim. \\ \hline
         AVERAGE & 0.72 & 0.55 & 0.66 & 0.53 & 0.84 & 0.75 \\ \hline
         STDEV & 0.049 & 0.09 & 0.10 & 0.07 & 0.10 & 0.06 \\ \hline
         MEDIAN & 0.71 & 0.54 & 0.63 & 0.54 & 0.82 & 0.75 \\ \hline
    \end{tabular}
    \caption{Similarity scores for similar and dissimilar (Dissim.) events using three different methods}
\label{tab:similar_dissimilar_pairs}
\end{table}

The results of the experiments are presented in Table
\ref{tab:similar_dissimilar_pairs} (w/o each data point). The average
similarity scores of both similar and dissimilar alert pairs are computed using
alert-description are 0.84 and 0.75, respectively. This shows that one cannot
rely only on the alert-description based similarity score, as it might result
in a lot of false positives. Although, precision will be high, but accuracy
will be low; these results indicate that the alert-description based similarity
score is not discriminative enough to discriminate similar pairs from
dissimilar pairs. On the other hand, the average similarity scores of both
similar and dissimilar alert pairs calculated using log-template are 0.72 and
0.55, respectively. It clearly indicates the discriminative power of
log-template based similarity scores, where it can clearly discriminate the
similar pairs from the dissimilar pairs. The entity based similarity scores
show similar results. That is, the average similarity score for similar and
dissimilar pairs are 0.66 and 0.53, respectively.

In order to establish the threshold value for alert grouping, i.e., when the
similarity between groups is less than a threshold then the
grouping should stop, we calculate the accuracy of log-template based alert grouping, entity 
based alert grouping, and alert-description based alert grouping for
different threshold values. Table \ref{tab:accuracy} shows that the 
precision of alert-description based alert grouping is the lowest, and the 
precision of log-template based alert grouping is the highest. The maximum 
precision of log-template alert grouping is \emph{0.89} when the threshold is 
set to \emph{0.65}---thus, this is used in \name.

\begin{table}
    \centering
    \scriptsize
    \begin{tabular}{|l|l|l|l|}
    \hline
         & Threshold=0.6 & 0.65 & 0.7 \\ \hline
        Alert-description & 0.34 & 0.38 & 0.48 \\ \hline
        Time \& Spatial & 0.72 & 0.75 & 0.79 \\ \hline
        Log-template & 0.82 & \textbf{0.89} & \textbf{0.89} \\ \hline
    \end{tabular}
    \caption{Precision of different event grouping models for varying values of threshold}
    \label{tab:accuracy}
    \vspace{-4mm}
\end{table}

\subsection{Feedback}
\label{subsec:eval_feedback}

\begin{figure}[t]
\center
\includegraphics[bb=0 0 100 100, trim=0 0 0 0, clip, width=0.4\textwidth]{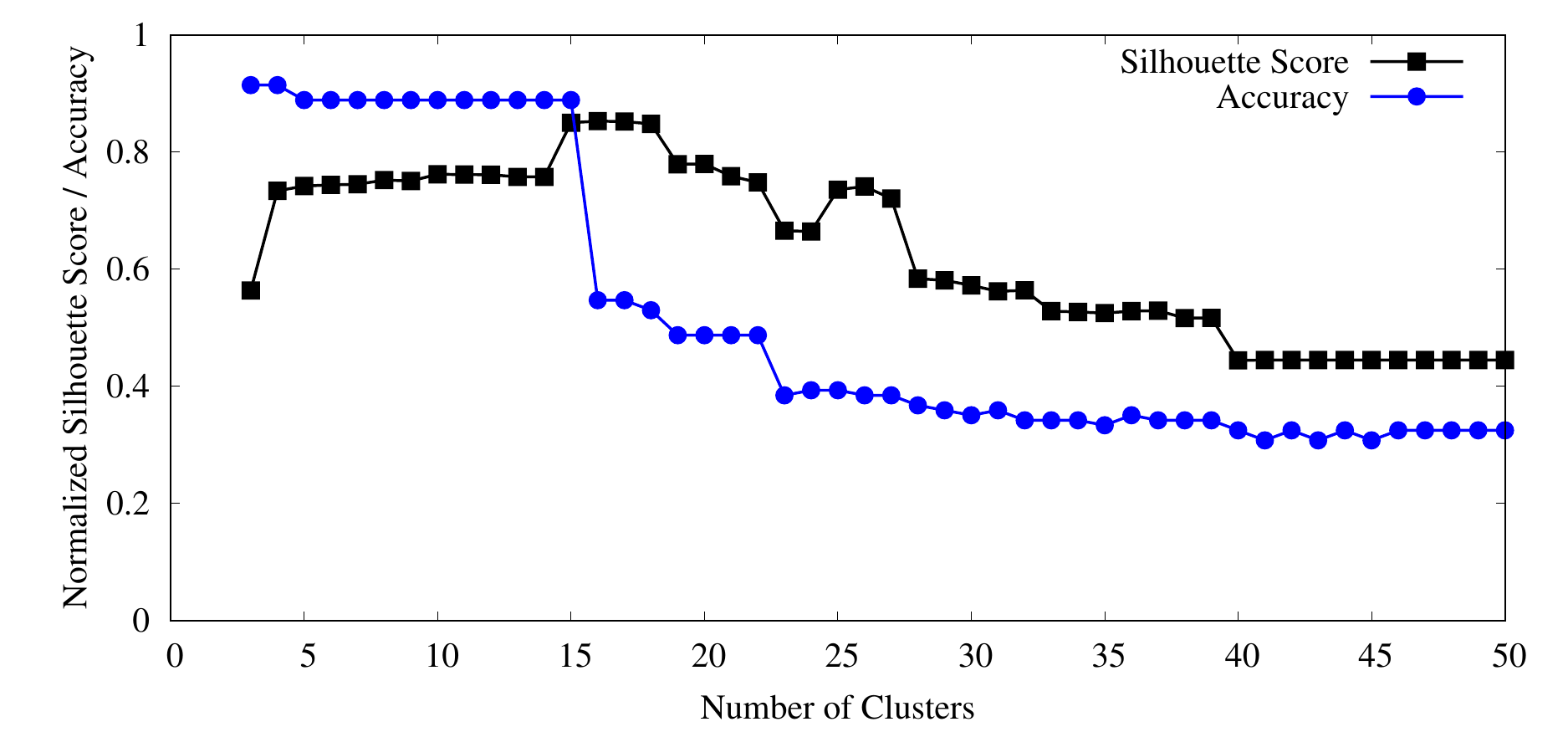}
\caption{Cluster consistency and accuracy for number of cluster}
\vspace{-3mm}
\label{fig:cluster}
\end{figure}

From the labeled data, we take the incorrect labels (6\% response rate) as
feedbacks.  As detailed in \S\ref{subsec:continuous_improvement}, \name
collects a simple feedback, yes/no from SREs, and uses them to perform
clustering \emph{split} and \emph{merge} operations to divide a cluster (group)
into multiple clusters or combine multiple clusters into one cluster. Both
cases help improve the correctness.  11,565 alerts have been used for training
word embeddings (size = 300) (\S\ref{subsec:learning_patterns}) and the word
embedding vectors are used as input to the clustering.
Figure~\ref{fig:cluster} illustrates the quality of clusters (i.e.,
consistency) using silhouette score and accuracy against the labeled data
(\S\ref{subsec:setup}).  15 clusters show the best consistency and accuracy.
From the labeled data (\S\ref{subsec:eval_event_correlation}), out of 68
incorrect labeled alerts, 46 alerts have been split and merged into another groups,
then become correct.  Also, 149 alerts (241 from single alert groups) have been
merged into other groups, then they become correct. Therefore, the precision
improves to $\frac{1020}{1134} = 0.9$ and F1 score becomes $0.95$.


\cut{
11,565 alerts texts
Document Embedding (Vector size = 300)
Birch clustering
Silhouette score to see how the clustering performs
T-SNE figure to plot clusters

Using the first set of ground-truth WA data (from SREs), we’d like to see how many clusters can be identified correctly to be grouped together
Note that clustering is done only with embedding data (no temporal info)

Results
1134 data points (events)
118 clusters (average 7.23 events, min = 2, max = 94)

Clusters capture similar events The event clustering can be used in both online
and offline with offline training model Finding the optimal number of clusters
needs careful investigation

The two can be a good indicator to decide how many clusters make sense
We’d like to provide good similarity for an object to its own cluster (cohesion) compared to other clusters (separation)
Also, overall the accuracy is high
}




\section{Conclusion}
\label{sec:conclusion}



We have introduced hybrid approaches used in the AIOps production systems, and
demonstrated that hybrid entity resolution and hybrid event correlation provide
better results than any single method.  Our experimental results clearly show
that combining template-based and dictionary-based approaches to entity
resolution achieves 93\% accuracy, while neither technique alone provides
sufficiently good results. We also show that consecutive application of
temporal and spacial methods together with association rule mining and
log-template grouping, help to improve performance. We conclude the
paper with description of methods used for feedback processing for improved
accuracy.  We have focused our initial work on data enrichment, correlation of
variety of data and fault localization methods as these two steps were
identified by SREs as most difficult in their troubleshooting process, we plan
to focus on root cause identification next.


%

\bibliographystyle{unsrt}
{\footnotesize
\bibliography{citations}}
\end{document}